\documentclass[aps,prd,reprint,amsmath,amssymb,floatfix,longbibliography]{revtex4-2}

\usepackage[T1]{fontenc}
\usepackage{lmodern}
\usepackage{microtype}
\usepackage{graphicx}
\usepackage{booktabs}
\usepackage{xcolor}
\usepackage[colorlinks=true,citecolor=blue!55!black,linkcolor=blue!55!black,urlcolor=blue!55!black]{hyperref}

\graphicspath{{figures/}}
\newcommand{\Msun}{\mathrm{M}_{\odot}}
\newcommand{\kpc}{\mathrm{kpc}}
\newcommand{\Gyr}{\mathrm{Gyr}}
\newcommand{\kms}{\mathrm{km\,s}^{-1}}
\newcommand{\sigmam}{\sigma/m}

\begin{document}

\title{Self-interacting dark matter delays bar growth in an isolated gas-rich galactic disk: torque-resolved numerical evidence}
\author{Dong-Biao Kang}
\affiliation{Intelligent Manufacturing College, Zhejiang Guangsha Vocational and Technical University of Construction, Jinhua 322100, China}
\affiliation{Institute of Theoretical Physics, Chinese Academy of Sciences, Beijing 100190, China}
\date{15 September 2026}

\begin{abstract}
We investigate how a dynamically responsive gas disk changes the influence of self-interacting dark matter (SIDM) on stellar-bar formation.  In an isolated Milky-Way-mass galaxy with fixed total baryonic mass, the gas-rich collisionless (CDM) model develops a persistent bar at $1.96\,\Gyr$, whereas the matched SIDM model with $\sigmam=1\,\mathrm{cm^2\,g^{-1}}$ does so at $2.69\,\Gyr$.  The final stellar $m=2$ amplitude is 0.437 in CDM and 0.340 in SIDM, a 22 per cent suppression; the corresponding gas-free controls remain weak over the same interval.  An independently sampled gas-rich pair preserves the direction of both differences.  Direct component-force measurements show that the SIDM stellar disk loses $2.82\times10^{11}\,\Msun\,\kpc\,\kms$ less angular momentum than its CDM counterpart.  The integrated torque contrast is dominated by the dark halo (approximately 74 per cent), with a substantial gas contribution (26 per cent), and is redistributed in radius and time rather than being a uniform reduction of halo coupling.  Conservation, force-accuracy, aperture, threshold, and independent-sampling checks support the ordering as a dynamical effect within this model.  The result demonstrates that a live gas component can change both the sign and timing of the early SIDM impact inferred from collisionless disks, providing a mechanism-based constraint for future hydrodynamic SIDM studies.
\end{abstract}

\keywords{Galactic bars, self-interacting dark matter, gas dynamics, numerical simulations}
\maketitle

\section{Introduction}

Stellar bars are long-lived non-axisymmetric structures that redistribute angular momentum, rearrange disk material, and contribute to the secular evolution of galaxies \citep{kormendy2004,sellwood2014}.  Their abundance depends on stellar mass, colour, and gas content \citep{erwin2018}, while their growth and slowdown are controlled by the exchange of angular momentum among the disk, halo, and any gaseous component.  The classical stability arguments of \citet{toomre1964} and \citet{ostriker1973} established that disk kinematics and halo support are central to bar formation.  Subsequent perturbative theory and simulations showed that a live halo can absorb angular momentum at resonances and thereby facilitate bar growth \citep{lyndenbell1972,tremaine1984,weinberg1985,athanassoula2002,athanassoula2003}.  The same coupling underlies the use of bar pattern speeds as probes of halo dynamical friction \citep{debattista2000}.

The microphysics of dark matter can alter this disk--halo exchange.  Self-interacting dark matter was introduced as a way to modify the inner structure of dark-matter halos while preserving the large-scale successes of cold dark matter \citep{spergel2000}.  Cosmological and idealized calculations have since established the formation of approximately isothermal central regions, the dependence of that response on cross section and scattering anisotropy, and the possibility of later gravothermal contraction \citep{vogelsberger2012,rocha2013,kaplinghat2016,robertson2017,balberg2002}.  Reviews by \citet{bullock2017} and \citet{tulin2018} summarize both the small-scale motivation and the allowed model space.  Because a bar communicates directly with halo orbits, it offers a complementary probe of SIDM that is sensitive to dynamics rather than to the instantaneous density profile alone.

Recent collisionless-disk simulations by \citet{dattathri2026} found that SIDM can accelerate bar formation and increase its amplitude.  Their interpretation is that scattering broadens bar--halo resonances and enhances the transfer of angular momentum from stars to dark matter.  That result is physically important, but the baryonic component in those calculations is collisionless.  Gas introduces pressure support, dissipation, radial transport, and an additional time-dependent gravitational torque.  Earlier simulations have shown that the influence of gas on bars is not described by a single monotonic rule: it depends on gas fraction, central concentration, halo structure, and numerical treatment \citep{villavargas2010,athanassoula2013}.  In particular, a gas-rich disk can change both bar growth and slowdown \citep{beane2023}.  It is therefore not guaranteed that the SIDM effect inferred from a collisionless stellar disk carries over unchanged to a hydrodynamic disk.

This paper reports a controlled first calculation of that interaction.  We compare CDM and SIDM at fixed total baryonic mass both without gas and with a 30 per cent gas fraction.  We then use a second independently sampled gas-rich pair, threshold and aperture variations, conservation diagnostics, and direct force recomputation to test whether the observed SIDM--CDM difference is dynamically meaningful.  Our central result is that, in the adopted warm adiabatic disk, SIDM delays the appearance of a persistent bar and reduces its amplitude over the first $3.91\,\Gyr$.  The corresponding stellar angular-momentum deficit is accounted for primarily by a change in the halo torque, with a smaller but non-negligible gas contribution.  Thus gas does not merely perturb the magnitude of the collisionless result; it changes the realized angular-momentum history enough to reverse the early-time ordering of the gas-rich CDM and SIDM bars in this model.

The evidential scope is important.  Matched evolution, an independent particle realization, conservation checks, diagnostic sensitivity tests, and direct force recomputation all support the reported ordering and exclude obvious integration or force-estimation artifacts.  We therefore present the result as a well-supported statement about this model family, while keeping its astrophysical interpretation deliberately model-specific.

\section{Galaxy models and numerical methods}

\subsection{Collisionless components}

The initial conditions describe an isolated disk embedded in a live, spherical halo.  The halo follows a smoothly truncated Navarro--Frenk--White profile \citep{navarro1997},
\begin{equation}
 \rho_{\rm h}(r)=\frac{\rho_s}{(r/r_s)(1+r/r_s)^2}
 \exp\left(-\frac{r}{r_{\rm cut}}\right),
\end{equation}
with total mass $M_{\rm h}=10^{12}\,\Msun$, scale radius $r_s=20\,\kpc$, and cutoff radius $r_{\rm cut}=200\,\kpc$.  The stellar disk has radial scale length $R_d=2.5\,\kpc$ and vertical scale height $h_z=0.3\,\kpc$.  Its density is proportional to $\exp(-R/R_d)\,\mathrm{sech}^2[z/(2h_z)]$.  There is no bulge in the models considered here.  The target minimum stellar Toomre parameter is $Q_{\min}=2.5$, with an exponentially declining radial velocity-dispersion profile.

The halo and stellar distribution functions are generated self-consistently with \textsc{Agama} \citep{vasiliev2019}.  For the gas-rich models, both collisionless components are regenerated in the converged combined halo--star--gas potential; gas is not added to a previously equilibrated collisionless realization.  This distinction avoids interpreting the immediate response to an inconsistent potential as bar physics.

\subsection{Gas disk and mass normalization}

The total baryonic mass is held fixed at
$M_{\rm bar}=3.88\times10^{10}\,\Msun$.  In the gas-free models it is entirely stellar.  In the gas-rich models, $M_\star=2.716\times10^{10}\,\Msun$ and $M_{\rm gas}=1.164\times10^{10}\,\Msun$, corresponding to $f_{\rm gas}=0.3$.  Stellar and gas particles have the same mass, $7.76\times10^4\,\Msun$.

The gas is a resolved warm layer with characteristic vertical scale $0.3\,\kpc$.  Its surface density is
\begin{equation}
 \Sigma_{\rm g}(R)=\Sigma_0
 \exp\left[-\frac{\sqrt{R^2+R_c^2}}{R_d}\right]T(R),
 \qquad R_c=0.3\,\kpc,
\end{equation}
where $T(R)=1$ inside $8R_d$, decreases with a continuously differentiable cosine taper over the next $2R_d$, and is zero outside $10R_d$.  The small central regularization prevents an unphysical negative value of the radial-equilibrium expression for $v_\phi^2$ while leaving the exponential profile essentially unchanged outside the central few hundred parsecs.

The vertical pressure profile is obtained from
\begin{equation}
 \frac{\partial P}{\partial z}=-\rho_{\rm g}\frac{\partial\Phi}{\partial z},
\end{equation}
with every radial column normalized to $\Sigma_{\rm g}(R)$.  The azimuthal velocity is then assigned from the same pressure and potential fields,
\begin{equation}
 v_\phi^2(R,z)=R\frac{\partial\Phi}{\partial R}
 +\frac{R}{\rho_{\rm g}}\frac{\partial P}{\partial R}.
\end{equation}
The gas obeys an ideal equation of state with $\gamma=5/3$.  Cooling, star formation, stellar feedback, and chemical evolution are omitted.  Consequently, this calculation isolates gravitational and adiabatic hydrodynamic coupling; it is not a model of a multiphase interstellar medium.

\subsection{Gravity, hydrodynamics, and self-interactions}
\label{sec:numerics}

The systems are evolved with \textsc{Gadget-4} \citep{springel2021}, using double-precision phase-space variables, a randomized domain centre, and hierarchical timesteps.  The tree gravity follows the class of algorithms introduced by \citet{barnes1986}; the adopted force-accuracy parameter is $6.25\times10^{-4}$.  All particle species use a gravitational softening length of $0.1\,\kpc$, and the maximum timestep is $10^{-3}$ in code units.  The hydrodynamic formulation descends from the coupled tree--SPH methodology of \citet{hernquist1989}; for context, modern mesh-free and SPH implementations and their numerical trade-offs are reviewed by \citet{hopkins2015}.  The time unit is $0.977792\,\Gyr$, so the final integration time $T=4$ corresponds to $3.91117\,\Gyr$.  The calculations use 16 MPI tasks; the domain decomposition is a performance choice and does not define the physical stochastic sequence.

SIDM is implemented as velocity-independent, isotropic, elastic scattering between dark-matter particles only.  The production estimator uses a compact normalized cubic-spline kernel with adaptive 64-neighbour support, symmetric endpoint coupling, and equal-mass dark-matter macro-particles.  Candidate pair probabilities are computed from the local kernel density, relative speed, cross section, and interaction interval.  An accepted event rotates the pair's relative velocity isotropically in the centre-of-mass frame, conserving pair momentum and kinetic energy.  Four probability subcycles are used, with an upper allowed pair probability of 0.1.  The CDM calculations use $\sigmam=0$; the SIDM calculations use $\sigmam=1\,\mathrm{cm^2\,g^{-1}}$.

The SIDM operator was tested before use in the galaxy calculations.  Uniform-density fixtures recover the analytic scattering rate at the sub-percent level for the adopted 64-neighbour estimator, momentum and pair energy are conserved to floating-point precision, and the macroscopic density response is stable across one, two, and four MPI ranks.  Resolution tests with 1024, 4096, and 16384 particles give event-rate differences of 4--7 per cent and median-density differences of 2--5 per cent relative to the largest fixture.  A separate implementation comparison using the public \textsc{Gizmo} framework \citep{hopkins2015} reproduces the primary structural response to within 1.1 per cent after correcting its interaction kick ordering; the raw event counts differ by about 13 per cent because the neighbour and event-selection conventions are not identical.  These are operator-level tests rather than a substitute for galaxy-scale resolution convergence.

\subsection{Simulation matrix}

Table~\ref{tab:models} lists the principal $2\times2$ experiment.  Each arm contains $1.5$ million particles.  Within a fixed gas fraction, the CDM and SIDM arms begin from the same phase-space realization and differ only in the scattering cross section.  Between gas fractions the equilibrium is regenerated because the component masses and pressure support differ.  All four arms are analysed at the same 14 output times between 0 and $3.91\,\Gyr$.

\begin{table*}[tbp]
\centering
\caption{Principal simulation matrix.  The total baryonic mass is fixed at $3.88\times10^{10}\,\Msun$.}
\label{tab:models}
\begin{tabular}{lrrrr}
\toprule
Model & $f_{\rm gas}$ & $\sigmam$ & $N_{\rm DM}$ & $(N_\star,N_{\rm gas})$\\
 & & ($\mathrm{cm^2\,g^{-1}}$) & & \\
\midrule
CDM--G0    & 0.0 & 0 & 1,000,000 & (500,000, 0)\\
SIDM1--G0  & 0.0 & 1 & 1,000,000 & (500,000, 0)\\
CDM--G30   & 0.3 & 0 & 1,000,000 & (350,000, 150,000)\\
SIDM1--G30 & 0.3 & 1 & 1,000,000 & (350,000, 150,000)\\
\bottomrule
\end{tabular}
\end{table*}

We also evolve an independently sampled gas-rich CDM/SIDM pair with 200,000 halo particles, 70,000 stellar particles, and 30,000 gas particles.  Its macroscopic mass model, softening, hydrodynamics, and integration horizon are unchanged.  Because both particle number and random realization differ from the principal pair, this calculation tests whether the direction of the SIDM effect survives a substantial change in sampling; it does not isolate a unique convergence rate.

\subsection{Bar and torque diagnostics}

For stellar particles in a specified cylindrical annulus, the normalized Fourier amplitude is
\begin{equation}
 A_m(R_1,R_2)=
 \frac{\left|\sum_j m_j\exp(i m\phi_j)\right|}{\sum_j m_j}.
\end{equation}
Our principal amplitude is $A_2$ within $0.5<R<10\,\kpc$; $0.5<R<5\,\kpc$ and $0.5<R<2.5\,\kpc$ provide aperture checks.  To distinguish a bar from an incoherent spiral response, the inner disk is divided into $0.25\,\kpc$ bins and we define
\begin{equation}
 C_2=\frac{|\sum_k Q_{2,k}|}{\sum_k|Q_{2,k}|},
 \qquad Q_{2,k}=\sum_{j\in k}m_j e^{2i\phi_j}.
\end{equation}
Only bins with at least 128 stellar particles are used.  A persistent-bar time is the first of three consecutive common outputs satisfying $A_2\geq0.15$ and $C_2\geq0.9$.  This operational definition is not a universal physical boundary; continuous amplitudes, radial profiles, and results for $A_2$ thresholds of 0.10 and 0.20 are reported alongside it.

The interaction contrast for any observable $X$ is
\begin{equation}
 \begin{split}
 \Delta_{\rm int}X={}&(X_{\rm SIDM,G30}-X_{\rm CDM,G30})\\
 &-(X_{\rm SIDM,G0}-X_{\rm CDM,G0}).
 \end{split}
\end{equation}
It separates the SIDM difference realized in the gas-rich pair from the contemporaneous difference in the gas-free pair.

To identify the dynamical origin of the gas-rich contrast, we recompute the gravitational acceleration on every one of the 350,000 stellar particles at all 14 common times.  Dark matter, gas, and stars are used as separate source sets with the same $0.1\,\kpc$ spline-softened gravity as the evolution calculation.  For source component $c$, the instantaneous torque is
\begin{equation}
 \tau_{z,c}(t)=\sum_i m_i\left[x_i a_{y,c}-y_i a_{x,c}\right].
\end{equation}
The primary force tree uses opening parameter $\theta=0.4$; endpoint forces are repeated with $\theta=0.25$.  Trapezoidal and Simpson time integrals test sensitivity to the sparse output cadence.  The directly measured change in stellar $L_z$ supplies an independent integral constraint.

\section{Results}

\subsection{Bar formation in the four principal models}

\begin{figure*}[tbp]
\centering
\includegraphics[width=0.96\linewidth]{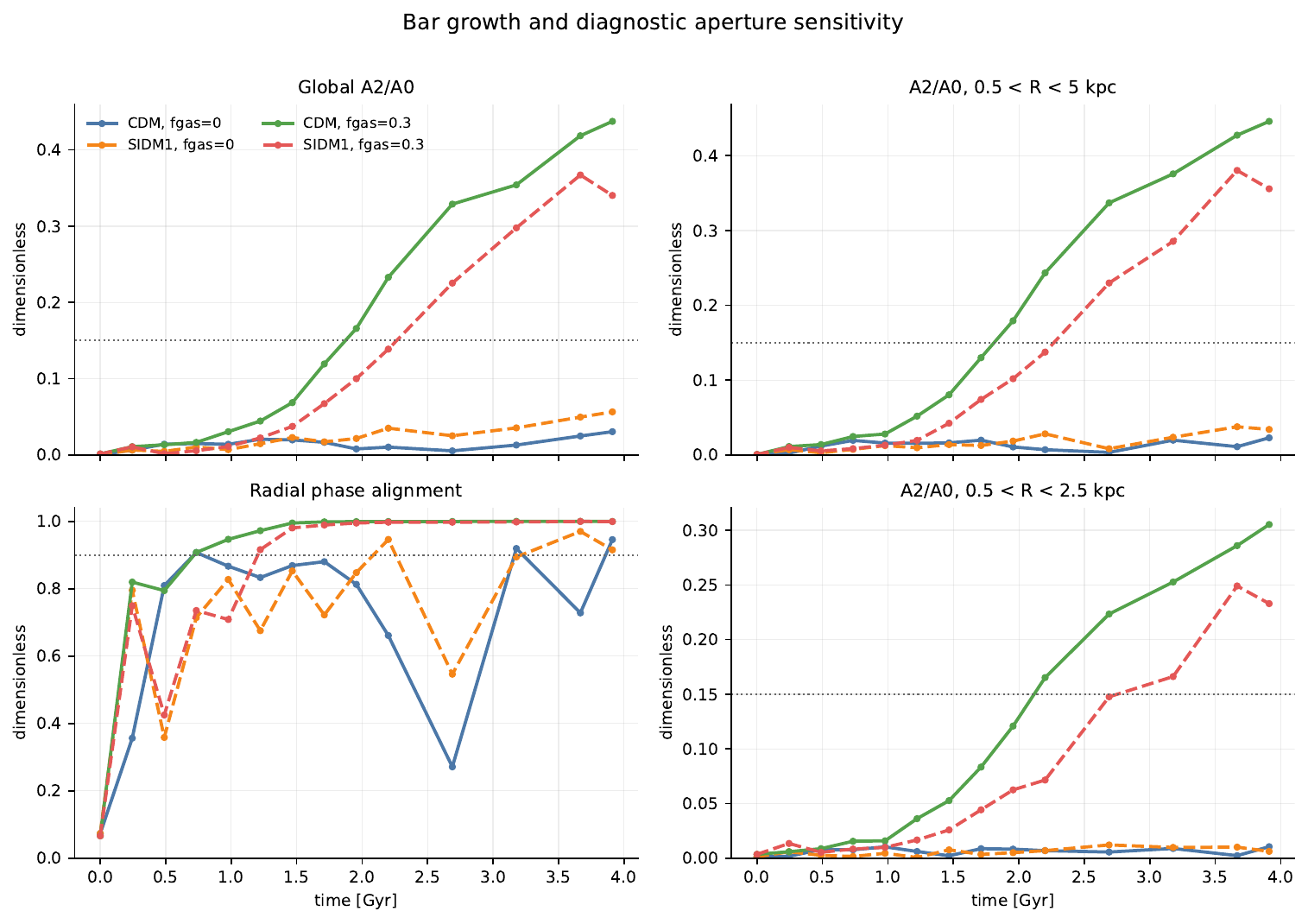}
\caption{Evolution of the principal four-model experiment.  The panels show the stellar $m=2$ amplitude within $0.5<R<10\,\kpc$ (top left), $0.5<R<5\,\kpc$ (top right), and $0.5<R<2.5\,\kpc$ (bottom right), together with the inner radial phase alignment (bottom left).  Gas-rich models form strong coherent bars; SIDM delays and suppresses that growth relative to CDM.}
\label{fig:growth}
\end{figure*}

Figure~\ref{fig:growth} shows the stellar $m=2$ amplitude in three radial apertures and the inner radial phase alignment.  The two gas-free disks remain weakly non-axisymmetric over the full interval: their endpoint $A_2$ values are 0.0305 in CDM and 0.0565 in SIDM.  Neither satisfies the persistent-bar definition in any tested aperture.  The comparison does not imply that these models never form bars; the collisionless fiducial model of \citet{dattathri2026}, for example, forms its CDM bar after approximately $11\,\Gyr$.  It establishes only that the gas-free controls remain in the low-amplitude phase over our common $3.91\,\Gyr$ horizon.

The gas-rich disks behave differently.  CDM--G30 first satisfies the persistent criterion at $1.956\,\Gyr$, whereas SIDM1--G30 does so at $2.689\,\Gyr$.  The output-defined delay is therefore $0.733\,\Gyr$.  The estimate is cadence limited: the immediately preceding intervals correspond to $0.244\,\Gyr$ for CDM and $0.489\,\Gyr$ for SIDM.  We consequently interpret the delay through the full growth curves and sensitivity tests, not through excessive precision in a single crossing time.

At $3.911\,\Gyr$, $A_2=0.4374$ in CDM--G30 and 0.3402 in SIDM1--G30.  The SIDM amplitude is lower by 0.0971, or 22.2 per cent relative to CDM.  The same ordering holds inside $5\,\kpc$, where $A_2=0.4459$ and 0.3557.  Both bars are strongly phase coherent, with $C_2=0.9996$ and 0.9993.  The endpoint difference is therefore a difference between two mature bars rather than contamination by an incoherent outer spiral.

\begin{figure*}[tbp]
\centering
\includegraphics[width=0.96\linewidth]{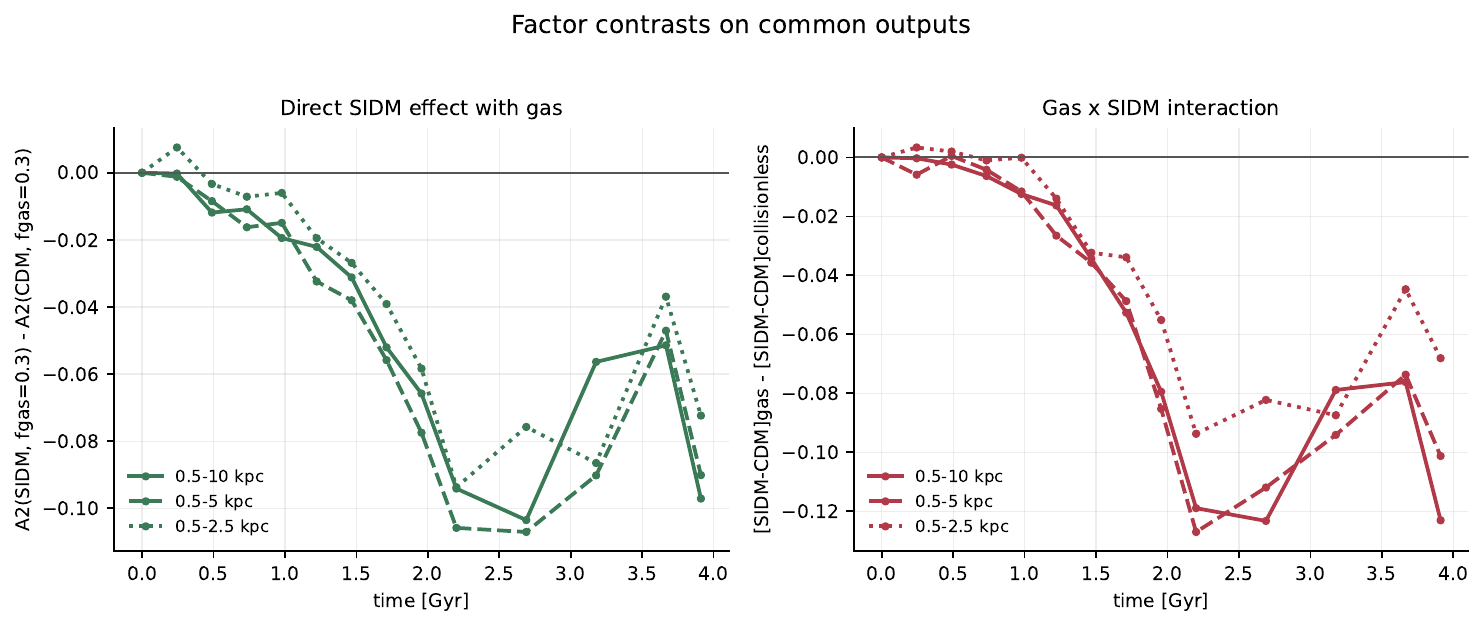}
\caption{Factor contrasts evaluated at common output times.  The gas-rich SIDM-minus-CDM amplitude becomes negative as the bar grows, while the corresponding gas-free difference is small and positive at the endpoint.  Their difference, $\Delta_{\rm int}A_2$, quantifies the non-additive SIDM--gas response.}
\label{fig:factor}
\end{figure*}

The endpoint interaction contrast is $\Delta_{\rm int}A_2=-0.1231$ (Fig.~\ref{fig:factor}).  Thus the weaker SIDM bar in the gas-rich pair is not explained by a general tendency of the SIDM realizations to have a smaller $m=2$ amplitude at this time: the gas-free SIDM model is instead 0.0260 higher than its CDM counterpart.  Because the gas-free bars are still weak, this interaction term should be read as an early nonlinear contrast rather than as a final asymptotic comparison of barred states.

\begin{table*}[tbp]
\centering
\caption{Principal bar and central-response measurements at $3.911\,\Gyr$.  The density ratio is relative to the initial mean dark-matter density within $1\,\kpc$.}
\label{tab:outcomes}
\begin{tabular}{lrrrr}
\toprule
Model & $A_2(0.5$--$10)$ & $A_2(0.5$--$5)$ & $C_2$ & $\bar\rho_{\rm DM}(<1)/\bar\rho_0$\\
\midrule
CDM--G0    & 0.0305 & 0.0229 & 0.9457 & 0.999\\
SIDM1--G0  & 0.0565 & 0.0341 & 0.9153 & 0.504\\
CDM--G30   & 0.4374 & 0.4459 & 0.9996 & 1.318\\
SIDM1--G30 & 0.3402 & 0.3557 & 0.9993 & 0.845\\
\bottomrule
\end{tabular}
\end{table*}

\subsection{Sensitivity to the bar definition and particle sampling}

Changing the amplitude threshold preserves the principal ordering.  For $A_2\geq0.10$, the CDM and SIDM gas-rich onset times are 1.711 and $1.956\,\Gyr$; for $A_2\geq0.20$, they are 2.200 and $2.689\,\Gyr$.  Varying the phase-alignment requirement from 0.8 to 0.95 does not change the onset times at the fiducial amplitude threshold.  At the final time SIDM has the smaller amplitude in all three apertures.  The inferred delay therefore does not depend on the particular choice $A_2=0.15$, although its numerical magnitude naturally changes with the definition.

The independently sampled 300,000-particle gas-rich pair provides a stronger test of stochastic sensitivity.  In the primary $0.5<R<10\,\kpc$ aperture, CDM crosses the persistent criterion at $1.222\,\Gyr$ and SIDM at $1.467\,\Gyr$, a delay of $0.244\,\Gyr$.  The endpoint growth relative to the initial amplitude is 0.2820 in CDM and 0.2764 in SIDM.  The difference is only $-0.0056$ in this broad aperture, but it is $-0.0522$ inside $5\,\kpc$ and $-0.0905$ inside $2.5\,\kpc$.  The onset delay is aperture dependent: zero inside $5\,\kpc$ and $0.733\,\Gyr$ inside $2.5\,\kpc$.  These variations demonstrate why a single crossing time is insufficient.  More importantly, the reduced pair does not reverse the qualitative result: SIDM never produces an earlier, stronger bar under the adopted primary comparison, and its inner endpoint amplitude is consistently smaller.

We regard this agreement as evidence that the sign is not peculiar to one particle realization.  The quantitative effect size remains sensitive to the sampling and aperture choices described above.

\subsection{Halo and gas response}

\begin{figure*}[tbp]
\centering
\includegraphics[width=0.96\linewidth]{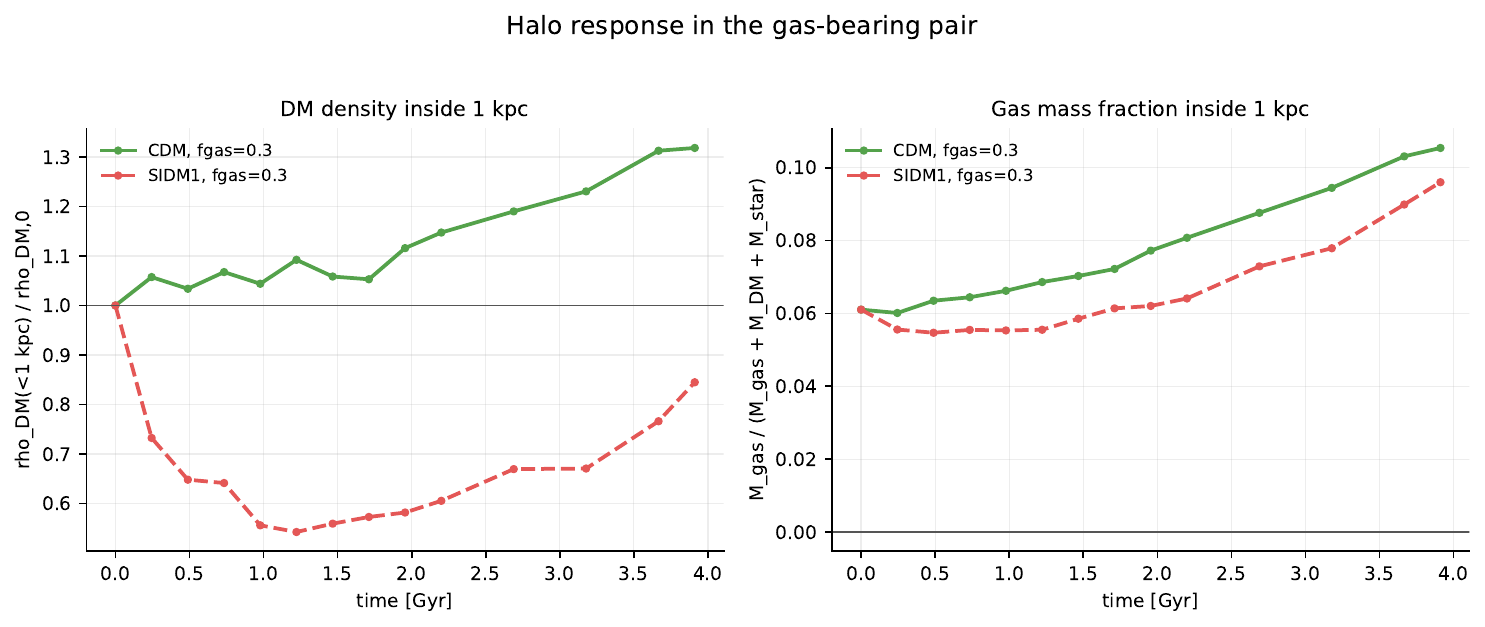}
\caption{Dark-matter and gas response in the principal models.  SIDM lowers the central halo density relative to CDM, while the gas-rich systems show additional contraction and radial redistribution.  The endpoint central gas fractions differ only modestly between the gas-rich pair.}
\label{fig:structure}
\end{figure*}

SIDM substantially alters the central halo in both gas fractions (Fig.~\ref{fig:structure}).  In the gas-free model the mean dark-matter density within $1\,\kpc$ falls to 0.504 of its initial value, compared with 0.999 in CDM.  In the gas-rich disks, the corresponding ratios are 0.845 in SIDM and 1.318 in CDM.  Thus the gas-rich SIDM halo is less centrally concentrated than its CDM counterpart even after baryonic rearrangement, but it is not simply the gas-free SIDM core transplanted into the gas model.

At the endpoint, 10.5 per cent of the gas lies inside $1\,\kpc$ in CDM--G30 and 9.6 per cent in SIDM1--G30.  The modest difference in central gas accumulation accompanies a much larger difference in stellar $A_2$.  Central density alone is therefore not a sufficient explanation of the bar ordering.  The relevant quantity is the coupled, time-dependent exchange of angular momentum.

\subsection{Angular-momentum evolution}

\begin{figure*}[tbp]
\centering
\includegraphics[width=0.96\linewidth]{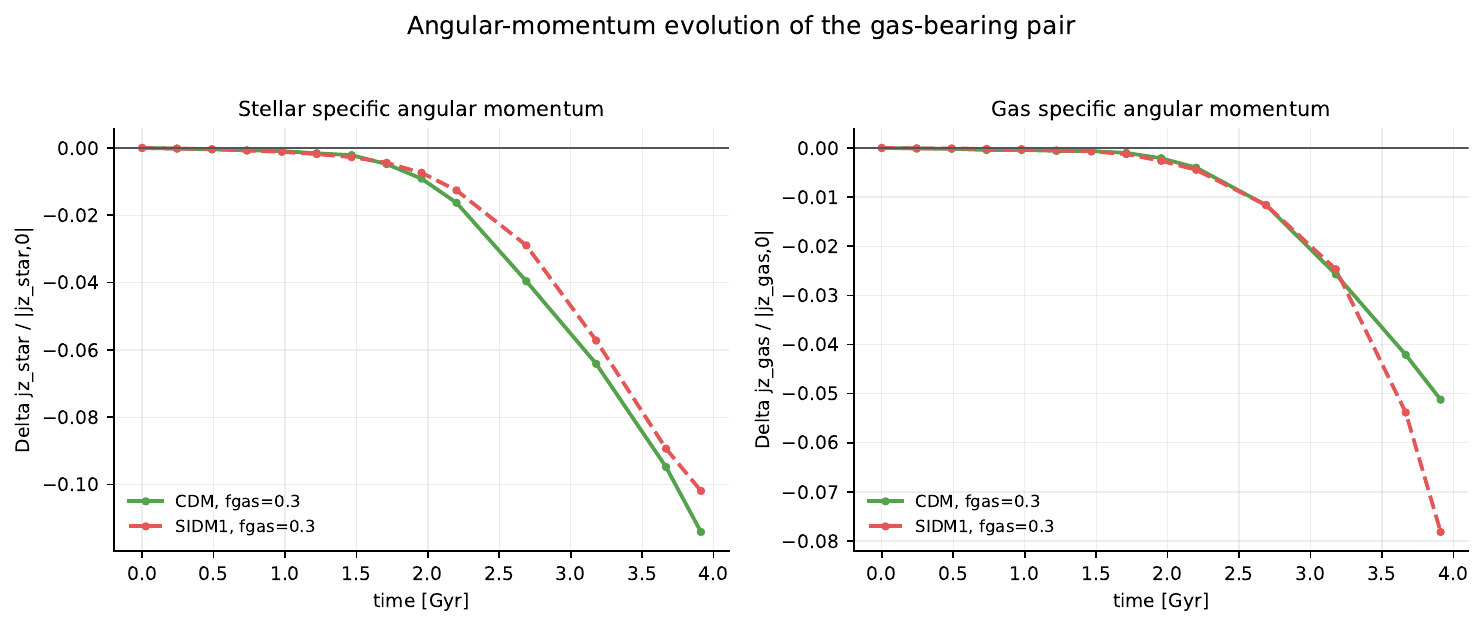}
\caption{Specific-angular-momentum evolution of the stellar and gas disks.  The gas-rich SIDM stellar disk loses less $L_z$ than the matched CDM disk over the full integration, consistent with its delayed and lower-amplitude bar.}
\label{fig:lz}
\end{figure*}

The stellar disk loses angular momentum as the bar grows (Fig.~\ref{fig:lz}).  Between the initial and final outputs the measured changes are
\begin{align}
 \Delta L_{z,\star}^{\rm CDM} &=-2.6400\times10^{12}\,
    \Msun\,\kpc\,\kms,\\
 \Delta L_{z,\star}^{\rm SIDM} &=-2.3578\times10^{12}\,
    \Msun\,\kpc\,\kms.
\end{align}
The SIDM stellar disk therefore loses $2.8216\times10^{11}\,\Msun\,\kpc\,\kms$, or 10.7 per cent, less angular momentum than the CDM disk.  This measured deficit has the sign required for the delayed and weaker SIDM bar.  It also motivates a direct decomposition of the forces rather than an interpretation based only on density profiles.

\section{Dynamical origin of the SIDM--CDM difference}

\subsection{Time-integrated component torques}

Table~\ref{tab:torque} gives the trapezoidal integrals of the recomputed torques.  The summed torque is more negative in CDM than in SIDM by $6.4801\times10^{11}\,\Msun\,\kpc\,\kms$.  Of this contrast, $4.7860\times10^{11}$ (73.9 per cent) is contributed by the dark-matter force and $1.6850\times10^{11}$ (26.0 per cent) by the gas force; the net stellar self-torque contributes only 0.14 per cent.  Simpson integration changes the absolute contrast to $9.83\times10^{11}\,\Msun\,\kpc\,\kms$ but preserves the hierarchy: 78.1 per cent halo and 22.0 per cent gas.  The percentages should therefore be interpreted descriptively, while the conclusion that the halo dominates and the gas is non-negligible is insensitive to the integration rule.

\begin{table*}[tbp]
\centering
\caption{Trapezoidal integrals of the gravitational torque on all stars in the gas-rich pair.  Units are $10^{12}\,\Msun\,\kpc\,\kms$.  The last column is SIDM minus CDM.}
\label{tab:torque}
\begin{tabular}{lrrr}
\toprule
Source & CDM--G30 & SIDM1--G30 & Difference\\
\midrule
Dark matter & $-2.9421$ & $-2.4635$ & $+0.4786$\\
Gas         & $-0.0107$ & $+0.1578$ & $+0.1685$\\
Stars       & $-0.0034$ & $-0.0025$ & $+0.0009$\\
All sources & $-2.9562$ & $-2.3082$ & $+0.6480$\\
\bottomrule
\end{tabular}
\end{table*}

The difference between the trapezoidal torque integral and the directly measured endpoint $\Delta L_{z,\star}$ is $-3.16\times10^{11}\,\Msun\,\kpc\,\kms$ for CDM (12.0 per cent of the measured change) and $+4.97\times10^{10}\,\Msun\,\kpc\,\kms$ for SIDM (2.1 per cent).  The larger CDM residual occurs during its rapid-growth interval, where the output cadence undersamples a strongly varying torque.  Simpson integration changes both residuals, confirming cadence sensitivity rather than a unique high-order correction.  We therefore use the direct endpoint $L_z$ difference for the total budget and the force integrals to assign its origin among components.

\subsection{Where and when the torque differs}

\begin{figure*}[tbp]
\centering
\includegraphics[width=0.96\linewidth]{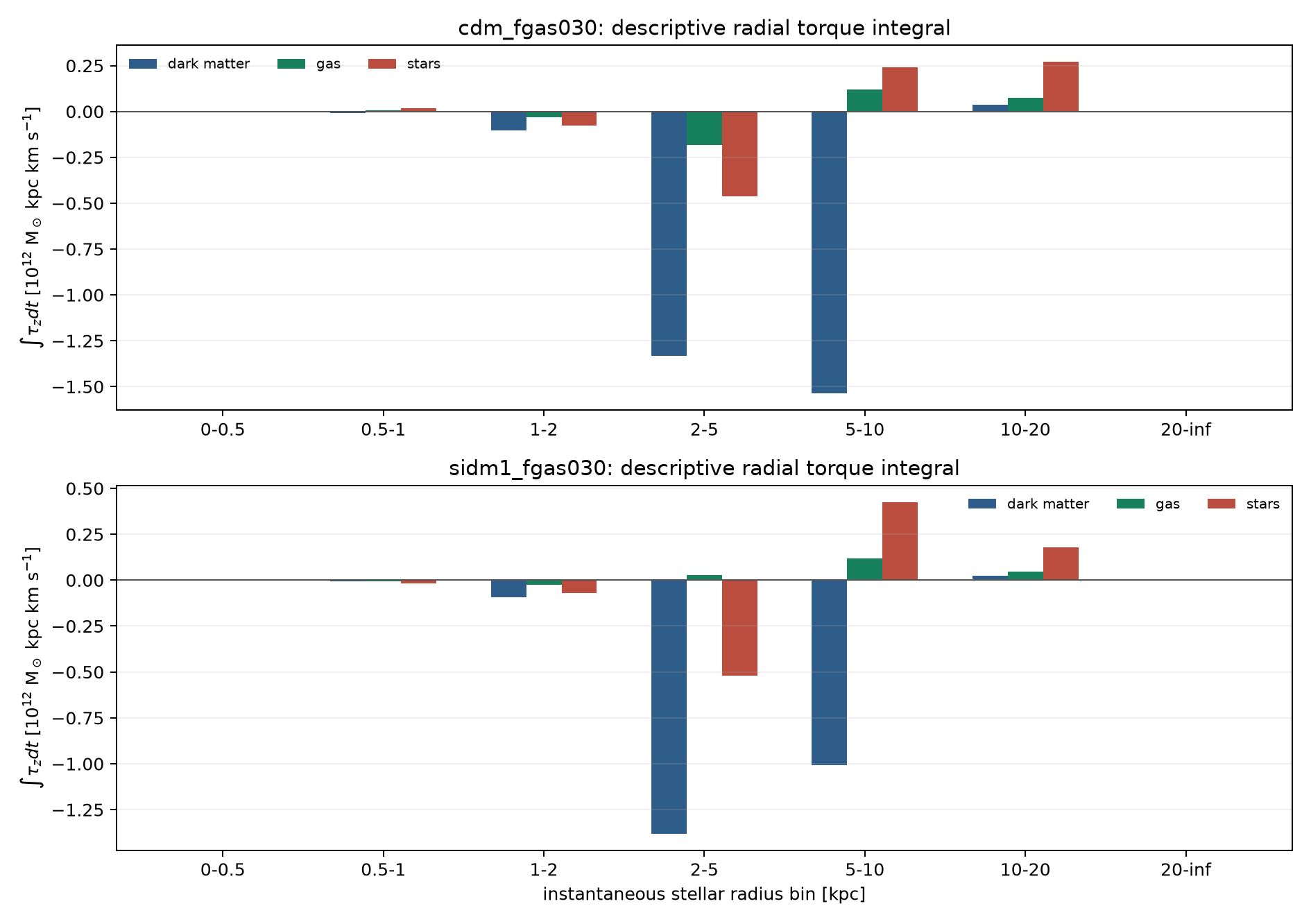}
\caption{Time-integrated gravitational torques on stars in instantaneous cylindrical-radius bins for CDM--G30 (top) and SIDM1--G30 (bottom).  The bins localize where each source force acts but are not Lagrangian control volumes because stars move between bins.  Comparing the panels shows that the largest halo contribution to the SIDM-minus-CDM contrast arises at $5<R<10\,\kpc$, while the largest gas contribution arises at $2<R<5\,\kpc$.}
\label{fig:radialtorque}
\end{figure*}

The SIDM-minus-CDM decomposition is not well described by a uniformly weaker halo torque.  Integrated over time, the dark-matter contrast is concentrated mainly among stars instantaneously located at $5<R<10\,\kpc$, where it is $+5.30\times10^{11}\,\Msun\,\kpc\,\kms$.  The $2<R<5\,\kpc$ halo contribution has the opposite sign, approximately $-5.02\times10^{10}\,\Msun\,\kpc\,\kms$.  The gas contrast is concentrated primarily at $2<R<5\,\kpc$, contributing $+2.11\times10^{11}\,\Msun\,\kpc\,\kms$ (Fig.~\ref{fig:radialtorque}).

The time dependence is equally important.  CDM exhibits a large negative halo torque earlier, around $T\simeq2.25$--2.75, as its bar enters rapid growth.  The SIDM halo torque becomes comparably negative later and is more negative at the final output, while the late gas torque in SIDM is positive.  These histories show that SIDM changes the timing and radial distribution of angular-momentum exchange.  The reduced early extraction of stellar $L_z$, rather than an absence of halo coupling, accounts for the delayed bar.

\begin{figure*}[tbp]
\centering
\includegraphics[width=0.96\linewidth]{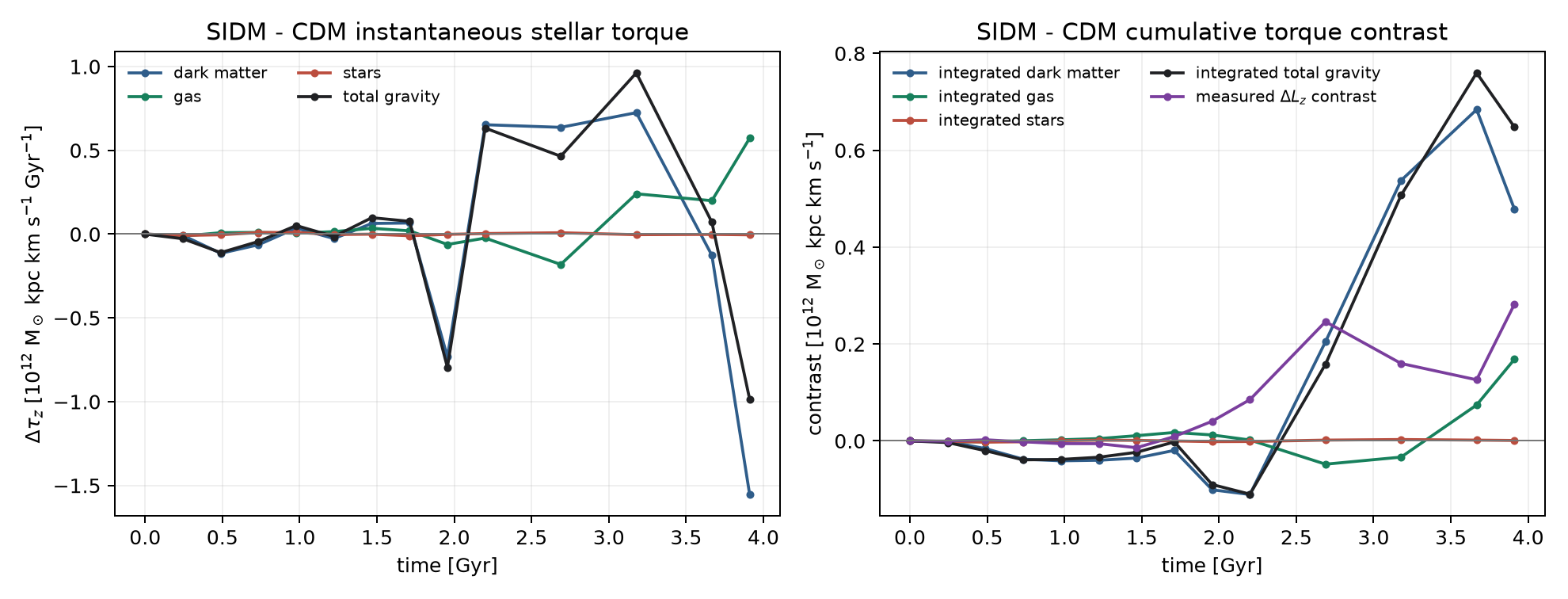}
\caption{SIDM-minus-CDM gravitational-torque contrast in the gas-rich pair, separated by source component and time.  Positive values mean that the SIDM torque is less negative (or more positive).  The halo dominates the integrated difference, but its sign and magnitude vary with time; gas supplies a smaller contribution with the same integrated sign.}
\label{fig:torquecontrast}
\end{figure*}

This constitutes direct dynamical evidence for the proposed mechanism at the level of resolved component forces: the altered halo and gas forces produce a smaller net loss of stellar angular momentum during the epoch when the CDM bar grows fastest.  It is not yet a resonance-space decomposition.  Establishing which orbital families and resonances carry the halo contribution would require reliable pattern speeds, actions, and frequency maps at a denser cadence, following the framework developed by \citet{lyndenbell1972}, \citet{tremaine1984}, and \citet{athanassoula2003}.

\section{Numerical reliability and limitations}

The scientific interpretation rests on several independent checks.

\begin{enumerate}
 \item \textit{Paired evolution.}  At each gas fraction, CDM and SIDM start from the same particle realization and use identical gravity, timesteps, hydrodynamics, softening, and output times.  Snapshot hashes, particle identifiers, times, and finite phase-space values were verified after transfer from the compute system.
 \item \textit{Integration accuracy.}  The maximum fractional total-energy drift among the principal calculations is below $2.4\times10^{-4}$.  This is far smaller than the 22 per cent endpoint difference in $A_2$ between the gas-rich models.
 \item \textit{Diagnostic stability.}  The gas-rich SIDM bar forms later for amplitude thresholds 0.10, 0.15, and 0.20, for phase-alignment requirements from 0.8 to 0.95, and its endpoint amplitude is smaller in all three radial apertures.
 \item \textit{Independent particle sampling.}  A second gas-rich pair with a different random realization and one-fifth the particle number retains the primary onset ordering and the inner-disk amplitude ordering.  The effect size varies, which is expected for nonlinear bar growth and is the main reason not to claim precision beyond the principal pair.
 \item \textit{Force accuracy.}  The force-decomposition implementation reproduces a four-particle analytic calculation with relative error $1.0\times10^{-8}$.  Tightening the tree opening parameter from 0.4 to 0.25 changes the endpoint all-source torque by 0.43 per cent in CDM and 1.77 per cent in SIDM.  The rms stellar tangential acceleration changes by at most 0.051 per cent, and the numerical net stellar self-torque is below 0.6 per cent of the rms total torque.
 \item \textit{Interaction and hydrodynamic separation.}  Isolated mixed-species tests confirm that gas particles are never selected as SIDM scattering endpoints.  In a matched zero/positive-cross-section gas test, the gas response is therefore mediated by gravity and hydrodynamics, as intended.
\end{enumerate}

These checks make a simple integration failure, force-tree error, threshold accident, or one particular random sample an implausible explanation of the observed ordering.  They do not remove all limitations.  A same-seed repeat at larger particle number would provide a formal resolution-convergence test for the nonlinear bar amplitudes and onset times.  The $0.1\,\kpc$ gravitational softening does not resolve the detailed structure of the innermost gas layer or parsec-scale SIDM scattering; our principal bar aperture begins at $0.5\,\kpc$, which reduces but does not eliminate sensitivity to the centre.  The gas is warm and adiabatic, so cooling, a multiphase medium, star formation, and feedback could change its central concentration and torque.  Only one non-zero cross section and one gas fraction are studied, and the sparse late-time outputs do not yield pattern speed or corotation radius with the accuracy needed for a resonance-by-resonance claim.

\section{Discussion}

\subsection{Relation to collisionless SIDM bar calculations}

The collisionless result of \citet{dattathri2026} and the gas-rich result reported here need not be contradictory.  In a collisionless disk, SIDM can broaden halo resonances and accelerate angular-momentum absorption.  Our direct force calculation shows that the halo remains the largest contributor to the SIDM--CDM torque difference, but the time sequence changes when a live gas disk is present.  During the early rapid-growth epoch, the SIDM gas-rich disk experiences less net negative torque; the gas contribution reinforces part of that difference.  By the final output, the instantaneous SIDM halo torque has caught up and can be more negative than in CDM.  A calculation that measured only the final torque would therefore miss the relevant history.

The central halo density also evolves differently.  SIDM creates a lower-density inner halo relative to CDM, consistent with the expected conductive response \citep{rocha2013,tulin2018}, but the bar contrast cannot be reduced to a monotonic mapping between central density and bar strength.  Bar growth depends on the phase-space distribution of halo orbits and on their coupling to the rotating perturbation, not only on enclosed mass.  The additional gas torque and the modified onset time further alter which orbital populations encounter the growing pattern.  This interpretation is compatible with the established sensitivity of bar evolution to live-halo structure and gas fraction \citep{athanassoula2002,villavargas2010,athanassoula2013}.

\subsection{Astrophysical implications}

The result identifies gas physics as a potentially decisive variable when using bars to constrain SIDM.  If an SIDM-induced acceleration measured in collisionless disks were applied directly to gas-rich galaxies, the inference could have the wrong sign during the first several gigayears.  This concern is especially relevant to dynamically young systems, where gas fractions and turbulent support are larger than in present-day massive spirals.  Conversely, our adiabatic model should not be treated as a prediction for an observed bar fraction.  Real gas cools, fragments, forms stars, and exchanges energy and momentum through feedback; each process can change the central mass distribution and the bar torque.

The most transferable conclusion is therefore mechanistic rather than demographic: in a coupled disk--halo--gas system, SIDM changes the radial and temporal angular-momentum pathway, and gas supplies a measurable part of the difference.  Any observational prediction must marginalize over this baryonic response.  Cross-section constraints based on bar incidence or strength should be constructed from matched hydrodynamic suites, not by adding a baryonic correction to collisionless SIDM calculations.

\section{Conclusions}

We have studied bar growth in an isolated Milky-Way-mass disk using a four-model comparison of $f_{\rm gas}=0$ and 0.3 with CDM and $\sigmam=1\,\mathrm{cm^2\,g^{-1}}$ SIDM.  The principal results are as follows.

\begin{enumerate}
 \item In the gas-rich pair, the CDM disk develops a persistent, phase-coherent bar at $1.96\,\Gyr$, while the SIDM disk reaches the same operational criterion at $2.69\,\Gyr$.  At $3.91\,\Gyr$, SIDM reduces the broad-aperture $m=2$ amplitude from 0.437 to 0.340, a 22 per cent difference.  The gas-free controls remain weak over this time and do not form comparable bars.
 \item The delay remains positive when the amplitude threshold is varied from 0.10 to 0.20, and the endpoint SIDM amplitude is lower in all tested apertures.  An independent 300,000-particle gas-rich pair preserves the primary onset ordering and the inner-amplitude ordering, although the effect size is sensitive to aperture and sampling.
 \item The SIDM stellar disk loses $2.82\times10^{11}\,\Msun\,\kpc\,\kms$ less angular momentum than the CDM disk.  A direct component-force calculation attributes most of the time-integrated torque contrast to the dark halo and a smaller, substantial fraction to gas.  The halo difference is redistributed across radius and time; it is not uniformly weak in SIDM.
 \item The physical ordering is supported by small energy errors, common-output and particle-identity checks, diagnostic sensitivity tests, an independent particle realization, validated SIDM operator tests, and force-tree refinement.  Within the tested model, these checks make a simple numerical artifact an unlikely explanation of the result.
\end{enumerate}

These results show that a live gaseous component can qualitatively alter the early influence of SIDM on bar formation.  A compact survey in $f_{\rm gas}$ and $\sigmam$ can determine where the sign changes.  Adding radiative cooling, star formation, and feedback is essential for connecting the mechanism to observed galaxies.  Denser temporal sampling during bar growth would permit robust pattern-speed and corotation measurements and, together with action--frequency analysis, would identify the resonances responsible for the halo contribution.  Those extensions can turn the model-specific dynamical result presented here into a predictive framework for using bar demographics and angular-momentum structure to constrain dark-matter self-interactions.

\section*{Data and software availability}

The source code for the initial-condition, SIDM, Gadget-4, and analysis components is publicly available at \url{https://github.com/billkang-x/sidm-diskbar}. The evolution calculations use \textsc{Gadget-4} \citep{springel2021}, whose modern design follows the public \textsc{Gadget} code family \citep{springel2005}; equilibrium distribution functions are constructed with \textsc{Agama} \citep{vasiliev2019}.

\begin{acknowledgments}

This work is supported by the grant No.2022KYQD-KDB from Zhejiang Guangsha Vocational and Technical University of Construction.

ChatGPT 5.6 (OpenAI) was used to assist with language polishing, code writing and debugging, and data analysis, including the preparation of analysis scripts and figures. The author directed this assistance by specifying the scientific objectives, simulation comparisons, diagnostic methods, and scope of interpretation, and by providing iterative instructions and corrections. The author manually reviewed the AI-assisted text, code, analyses, and interpretations and checked the reported results for consistency with the simulation data and the numerical validation and sensitivity diagnostics described in this paper. A record of AI use is retained with the project documentation. The author assumes full responsibility for the entire content of the manuscript, including its accuracy, integrity, and conclusions. The AI system is not an author.
\end{acknowledgments}

\appendix
\section{Additional torque diagnostics}

\begin{figure*}[tbp]
\centering
\includegraphics[width=0.96\linewidth]{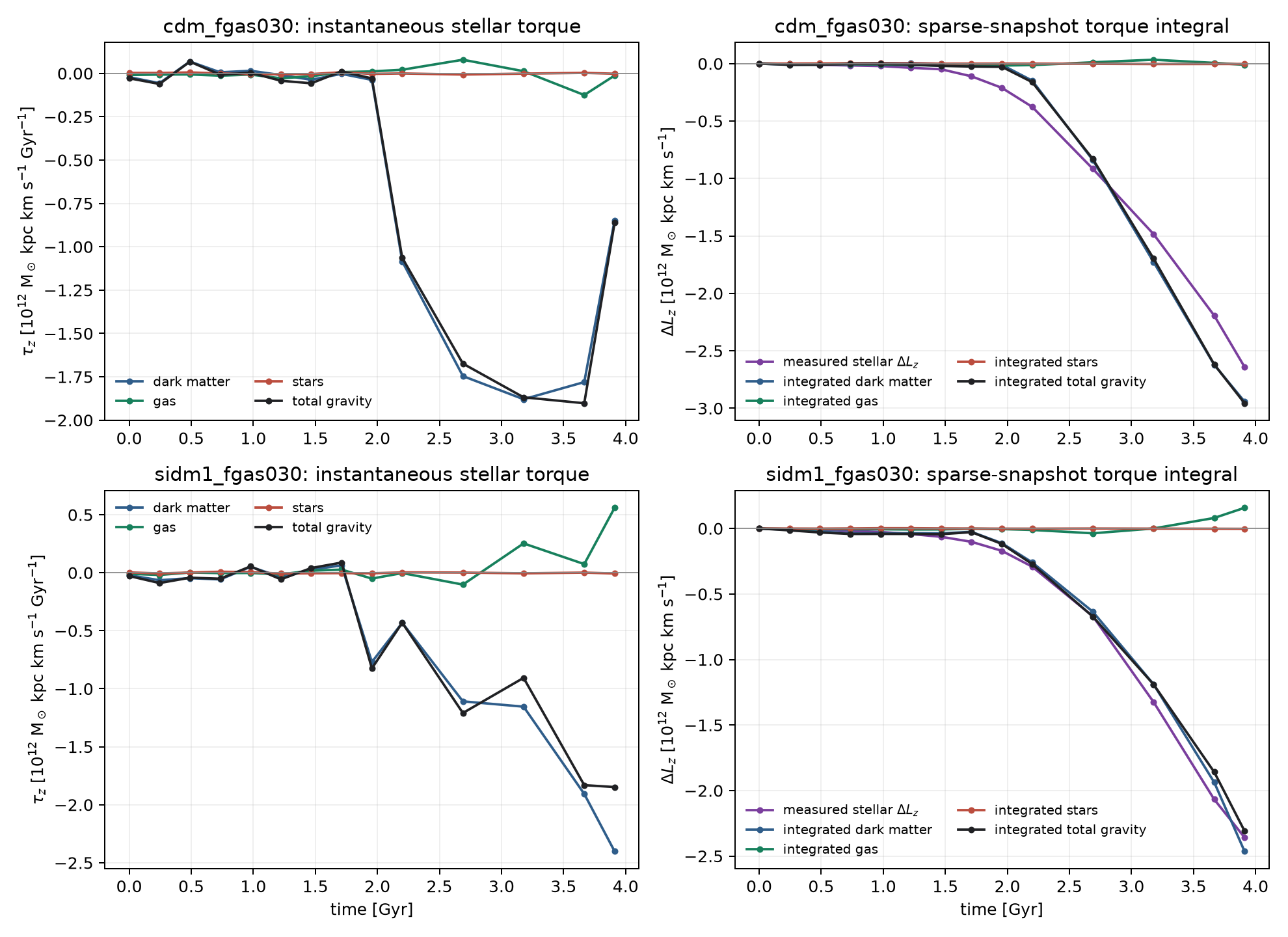}
\caption{Instantaneous and cumulative gravitational torques on the stellar disk, separated into dark-matter, gas, and stellar source components.  The vertical sampling is fixed by the 14 archived common outputs; cumulative values should therefore be compared with the directly measured endpoint $L_z$ changes quoted in the text.}
\label{fig:torquetimeseries}
\end{figure*}

Figure~\ref{fig:torquetimeseries} displays the component torques and their cumulative trapezoidal integrals.  It makes explicit that the principal contrast is accumulated during the period of rapid CDM bar growth rather than at a single endpoint.  The near-zero net stellar self-torque is an internal consistency check: pairwise stellar forces should not change the total stellar angular momentum except through finite force-estimation errors.

\section{Scope of the numerical-method comparison}

The SIDM implementation is stochastic, so exact event-by-event agreement across independent codes is neither expected nor required.  The relevant checks are conservation for every accepted pair, recovery of the ensemble scattering rate, convergence of density and velocity-distribution responses, and invariance of macroscopic results to MPI decomposition.  Our tests satisfy these conditions at the accuracy quoted in Sec.~\ref{sec:numerics}.  The public comparison implementation uses different neighbourhood and proposal conventions, which explains why raw accepted-event counts differ more than the primary structural response.  This distinction is important when interpreting cross-code validation: agreement of observables is stronger evidence than agreement of one implementation-dependent counter.

The gravitational part of the force calculation is independently controlled.  The same spline softening is used in evolution and post-processing, and the tree-refinement changes are much smaller than the component-torque contrast.  The method builds on standard hierarchical gravity and particle hydrodynamics \citep{barnes1986,hernquist1989,springel2005}, while the force recomputation is tailored to the component separation required here.

\bibliographystyle{apsrev4-2}
\bibliography{references}

\end{document}